# Analytical Model of Compact Star with a new version of Modified Chaplygin Equation of State

Manuel Malaver[1] and Rajan Iyer[2]

[1]Maritime University of the Caribbean, Department of Basic Sciences, Catia la Mar, Venezuela.
Email: **mmf.umc@gmail.com**

[2]Environmental Materials Theoretical Physicist, Department of Physical Mathematics Sciences Engineering Project Technologies, Engineeringinc International Operational Teknet Earth Global, Tempe, Arizona, United States of America
Email: engginc@msn.com

**Abstract:** In this paper we found a new model for compact star with anisotropic matter distribution considering the new version of Chaplygin fluid equation of state of Errehymy and Daoud (2021). We specify the particular form of the metric potential proposed for Thirukanesh and Ragel (2012) and generalized for Malaver (2014) in order to integrate the Einstein`s field equations. The obtained model satisfies all physical properties expected in a realistic star. The radial pressure, energy density, metric coefficients, anisotropy and mass are well defined and are regular in the stellar interior. The results of this research can be useful in the development and description of new models of compact structures.

## 1. Introduction

The general theory of relativity has allowed a better understanding of the structure of the universe [1]. The search for and finding of exact solutions of Einstein's field equations is an important area of research in the study of stellar structure, due to significant progress that has been made to model the interior of compact objects [2,3]. Within this context it is appropriate to mention the findings of Delgaty and Lake [4] who constructed several analytic solutions that can describe realistic stellar configurations and satisfy all the necessary conditions to be physically acceptable. These exact solutions have also made it possible the way to study cosmic censorship and analyze the formation of naked singularities [3].

In the development of the first stellar models it is important to mention the pioneering research of Schwarzschild [5], Tolman [6], Oppenheimer and Volkoff [7] and Chandrasekhar [8]. Schwarzschild [5] obtained interior solutions that allows describing a star with uniform density, Tolman [6] generated new solutions for static spheres of fluid, Oppenheimer, and Volkoff [7] studied the gravitational equilibrium of neutron stars using

Tolman's solutions and Chandrasekhar [8] produced new models of white dwarfs in presence of relativistic effects.

A great number of exact models from the Einstein-Maxwell field equations have been generated by Gupta and Maurya [9], Kiess [10], Mafa Takisa and Maharaj [11], Malaver and Kasmaei [12], Malaver [13,14], Ivanov [15] and Sunzu et al [16]. For the construction of these models, several forms of equations of state can be considered [17]. Komathiraj and Maharaj [18], Malaver [19], Bombaci [20], Thirukkanesh and Maharaj [21], Dey et al. [22] and Usov [23] assume linear equation of state for quark stars. Feroze and Siddiqui [24] considered a quadratic equation of state for the matter distribution and specified particular forms for the gravitational potential and electric field intensity. MafaTakisa and Maharaj [11] obtained new exact solutions to the Einstein-Maxwell system of equations with a polytropic equation of state. Thirukkanesh and Ragel [25] have obtained particular models of anisotropic fluids with polytropic equation of state which are consistent with the reported experimental observations. Malaver [26] generated new exact solutions to the Einstein-Maxwell system considering Van der Waals modified equation of state with polytropic exponent. Malaver and Kasmaei proposed a new model of compact star with charged anisotropic matter using a cosmological Chaplygin fluid [27]. Bertolami and Páramos [28] have studied general features of a spherically symmetric object described through the generalized Chaplygin fluid equation of state. Tello-Ortiz et al. [29] found an anisotropic fluid sphere solution of the Einstein-Maxwell field equations with a modified Chaplygin equation of state. Errehymy and Daoud [30] considered a distribution of anisotropic matter with a new version of modified Chaplygin fluid equation of state dependent on two parameters $\alpha$ and $\beta$. Prasad et al.[31] proposed a new model of an anisotropic compact star which admits the Chaplygin equation of state considering the metric potential of Buchdahl [32]. More recently, Malaver et. al [33] obtained new solutions of Einstein's field equations in a Buchdahl spacetime considering a nonlinear electromagnetic field.

General relativity also allows the analysis, through Einstein's gravity theory, of different cosmological scenarios as the existence of dark energy, dark matter, Phantom and Quintessence fields that were introduced to explain the accelerated expansion of the universe [34,35]. The Chaplygin fluid model whose equation of state $P = -\frac{B}{\rho}$ where $P$ is the pressure, $\rho$ the energy-density and $B$ a positive constant, has been considered an alternative to the Phantom and Quintessence fields [34-37] and is well-consistent with various classes of observational tests such as supernovae data [37], gravitational lensing [38,39], gamma ray bursts [40] and cosmic microwave background radiation [41].

The analysis of compact objects with anisotropic matter distribution is very important, because that the anisotropy plays a significant role in the studies of relativistic spheres of fluid [42-55]. Anisotropy is defined as $\Delta = p_t - p_r$ where $p_r$ is the radial pressure and $p_t$ is the tangential pressure. The existence of solid core, presence of type 3A superfluid [56], magnetic field, phase transitions, a pion condensation and electric field [23] are most important reasonable facts that explain the presence of tangential pressures within a star. Many astrophysical objects as X-ray pulsar, Her X-1, 4*U*1820-30 and SAXJ1804.4-3658 have anisotropic pressures. Bowers and Liang [54] include in the equation of hydrostatic equilibrium the case of local anisotropy. Bhar et al. [57] have studied the behavior of relativistic objects with locally anisotropic matter distribution considering the Tolman VII form for the gravitational potential with a linear relation between the energy density and the radial pressure. Malaver [58-59], Feroze and Siddiqui [24,60] and Sunzu et al.[16] obtained solutions of the Einstein-Maxwell field equations for charged spherically symmetric space-time by assuming anisotropic pressure.

For compact star with quark matter and other stellar models in general relativity [61, 62], we would recognize that energetic stars would undergo vortex action fields, that have been modeled by breakthrough formalism examining quantum fields point model algorithmically gaging to electromagnetic fields provided stringmetrics that are associated quantum to mesoscopic to astrophysics [63-67]. We have also incorporated hod Plenum PDP circuit assemblages to identify fundamental mechanism generating sustainable energy at quantum to astrophysical levels. [63]. Our analysis has also further shone light on the vacuum friction that tells of tired light or the inertial matter linking to non-inertial vacuum. Star systems essentially balance gas clouds such as Chaplygin equation of state to hydrogenous energetics within vacuum multiverse [23, 34, 35, 56, 61, 62]. In this respect, signal/noise density matrix will play key role in determining permutative combinatorial entanglement with prime factorized magic square symmetries universal mechanism versus dissociative decoherence entropic system environment. We believe that magic square symmetry prime factorization will eventually differentiate among inertial, charged, and neutral matter [66]. Mesoscopic parameters temperature and pressure have been analyzed exemplifying observables gage fields and have anisotropy induced by magic square prime factorization mechanisms, theoretically logically comprehensible algebra gage physics with unitary matrix properties [65, 66]. Hence critical signal/noise density matrix values may input to astrophysics metrics that are further quantified here in this paper of analytical model of compact, anisotropic, strange stars physical features associating the matter, radial pressure, density, anisotropy, gravitational potential, and energy density Schwarzschild-Einstein-Maxwell metrics.

In this paper, we generated a model of compact object with a new version of Chaplygin equation of state proposed for Errehymy and Daoud [30] that has the $p_r = H\rho^\alpha - K\rho^{-\beta}$ where *H, K, α* and *β* are constants. We have used the Thirukkanesh-Ragel-Malaver ansatz [25,49,61] that is nonsingular, continuous and well behaved in the interior of the star and has been obtained a new class of static spherically symmetrical model for a anisotropic matter distribution. It is expected that the solution obtained in this work can be applied in the description and the study of internal structure of compact objects. The paper is organized as follows: In the section "Einstein-Maxwell Equations", we present Einstein-Maxwell field equations. In the section "New Class of Anisotropic Strange Stars", we make a particular choice for gravitational potential Z(x) and generated new models for anisotropic matter. In the section "Physical Requirements for the New Model", physical acceptability conditions are discussed. The physical properties and physical validity of these new solutions are analyzed in the "Physical Analysis" section. The conclusions of the results obtained are shown in the "Conclusion" section.

## 2. Einstein-Maxwell Field Equations

We consider a spherically symmetric, static and homogeneous spacetime. In Schwarzschild coordinates the metric is given by

$$ds^2 = -e^{2\nu(r)}dt^2 + e^{2\lambda(r)}dr^2 + r^2(d\theta^2 + \sin^2\theta d\varphi^2)  \quad (1)$$

where $\nu(r)$ and $\lambda(r)$ are two arbitrary functions.

The Einstein field equations for the charged anisotropic matter are given by

$$\frac{1}{r^2}(1-e^{-2\lambda}) + \frac{2\lambda'}{r}e^{-2\lambda} = \rho \quad (2)$$

$$-\frac{1}{r^2}(1-e^{-2\lambda}) + \frac{2\nu'}{r}e^{-2\lambda} = p_r \quad (3)$$

$$e^{-2\lambda}\left(\nu'' + \nu'^2 + \frac{\nu'}{r} - \nu'\lambda' - \frac{\lambda'}{r}\right) = p_t \quad (4)$$

where $\rho$ is the energy density and $p_r$ is the radial pressure, $p_t$ is the tangential pressure and primes denote differentiations with respect to r. Using the transformations, $x = cr^2$, $Z(x) = e^{-2\lambda(r)}$ and $A^2 y^2(x) = e^{2\nu(r)}$ with arbitrary constants A and c>0, suggested by Durgapal and Bannerji [62], the Einstein field equations can be written as

$$\frac{1-Z}{x} - 2\dot{Z} = \frac{\rho}{c} \tag{5}$$

$$4Z\frac{\dot{y}}{y} - \frac{1-Z}{x} = \frac{p_r}{c} \tag{6}$$

$$4xZ\frac{\ddot{y}}{y} + (4Z + 2x\dot{Z})\frac{\dot{y}}{y} + \dot{Z} = \frac{p_t}{c} \tag{7}$$

$$p_t = p_r + \Delta \tag{8}$$

$$\frac{\Delta}{c} = 4xZ\frac{\ddot{y}}{y} + \dot{Z}\left(1 + 2x\frac{\dot{y}}{y}\right) + \frac{1-Z}{x} \tag{9}$$

$\Delta = p_t - p_r$ is the anisotropic factor and dots denote differentiation with respect to x. With the transformations of [62], the mass within a radius r of the sphere takes the form

$$M(x) = \frac{1}{4c^{3/2}} \int_0^x \sqrt{x}\rho(x)dx \tag{10}$$

In this paper, we assume the following equation of state

$$p_r = H\rho^\alpha - \frac{K}{\rho^\beta} \qquad \text{with } 0 \leq \beta \leq 1 \tag{11}$$

### 3. New Class of Anisotropic Stars

In this research, we have chosen the Thirukanesh-Ragel-Malaver ansatz [25,49,61] as metric potential which has the form $Z(x) = (1-ax)^n$, where $a$ is a real constant and $n$ is an adjustable parameter. This potential is regular at the stellar center and well behaved in the interior of the sphere. We have considered the particular cases $n=1, 2, 3$ but for $n=1$ the radial pressure and energy density are constant in the center and the surface of the sphere, for what it is not a physically acceptable solution [49].

For the case $n=2$, substituting $Z(x)$ in equation (5) we have for the energy density

$$\rho = ac(6-5ax) \qquad (12)$$

Using eq. (12) in eq.(11) and taking into account that for all the cases considered in this paper $\alpha=2$ and $\beta=1$, the radial pressure can be written in the form

$$p_r = Ha^2c^2(6-5ax)^2 - \frac{K}{ac(6-5ax)} \qquad (13)$$

and for the mass function we obtain

$$M(x) = \frac{ax^{3/2}(2-ax)}{2\sqrt{c}} \qquad (14)$$

Replacing eq. (13) and $Z(x)$, the eq.(6) becomes

$$\frac{\dot{y}}{y} = \frac{Ha^2c(6-5ax)^2}{4(1-ax)^2} - \frac{K}{4ac^2(1-ax)^2(6-5ax)} + \frac{2a-a^2x}{4(1-ax)^2} \qquad (15)$$

Integrating eq. (15), we have

$$y(x) = c_1(5ax-6)^{A*}(ax-1)^B e^{\frac{Cx^2+Dx+E}{4(ax-1)}} \qquad (16)$$

where $c_1$ is the constant within integration procedures.

For convenience we have let

$$A* = \frac{5K}{4c^2a^2} \qquad (17)$$

$$B = -\frac{5Hac}{2} - \frac{1}{4} - \frac{5K}{4c^2a^2} \qquad (18)$$

$$C = 25Ha^3c \tag{19}$$

$$D = -25Ha^2c \tag{20}$$

$$E = -Hac - 1 + \frac{K}{c^2 a^2} \tag{21}$$

For the metric functions $e^{2\lambda}$, $e^{2\nu}$ we have

$$e^{2\lambda} = \frac{1}{(1-ax)^2} \tag{22}$$

$$e^{2\nu} = A^2 c_1^2 (5ax-6)^{2A^*} (ax-1)^{2B} e^{\frac{Cx^2+Dx+E}{2(ax-1)}} \tag{23}$$

and the anisotropy $\Delta$ can be written as

$$\frac{\Delta}{c} = 4x(1-ax)^2 \frac{\ddot{y}}{y} - 2a(1-ax)\left(1 + 2x\frac{\dot{y}}{y}\right) + 2a - a^2 x \tag{24}$$

With n=3, the expression for the energy density is

$$\rho = ac(9 - 15ax + 7a^2 x^2) \tag{25}$$

replacing eq. (25) in eq. (11) and with α=2 and β=1, we have for the radial pressure

$$p_r = Ha^2 c^2 (9 - 15ax + 7a^2 x^2)^2 - \frac{K}{ac(9 - 15ax + 7a^2 x^2)} \tag{26}$$

and the mass function is

$$M(x) = \frac{ax^{3/2}(3 - 3ax + a^2 x^2)}{2\sqrt{c}} \tag{27}$$

Substituting eq. (26) and $Z(x)$ *in* eq. (6) we obtain

$$\frac{\dot{y}}{y} = \frac{Ha^2 c(9 - 15ax + 7a^2 x^2)^2}{4(1-ax)^3} - \frac{K}{4ac^2(1-ax)^3(9 - 15ax + 7a^2 x^2)} + \frac{3a - 3a^2 x + a^3 x^2}{4(1-ax)^3} \tag{28}$$

Integrating eq. (28), we have

$$y(x) = c_2 (7a^2 x^2 - 15ax + 9)^F (ax - 1)^G e^{-\frac{Ix^4 + Jx^3 + Lx^2 + Mx + N + (40a^2 x^2 - 80ax + 40)K\sqrt{3}\arctan\left[\frac{1}{9}(14ax - 15)\sqrt{3}\right]}{72a^2 c^2 (ax-1)^2}} \tag{29}$$

Again for convenience we have let

$$F = \frac{3K}{4c^2 a^2} \tag{30}$$

$$G = -\frac{15Ha^3 c^3 + 6K + a^2 c^2}{4c^2 a^2} \tag{31}$$

$$I = 441 Ha^7 c^3 \tag{32}$$

$$J = -2016 Ha^6 c^3 \tag{33}$$

$$L = 2709 Ha^5 c^3 \tag{34}$$

$$M = 18aK - 1098 Ha^4 c^3 + 18a^3 c^2$$

(35)

$$N = -45a^3 c^3 - 27a^2 c^2 - 9K \tag{36}$$

And for the metric functions $e^{2\lambda}$, $e^{2\nu}$ and anisotropy $\Delta$ we have

$$e^{2\lambda} = \frac{1}{(1-ax)^3} \tag{37}$$

$$e^{2\nu} = A^2 c_2^2 \left(7a^2x^2 - 15ax + 9\right)^{2F} (ax-1)^{2G} e^{-\frac{Ix^4 + Jx^3 + Lx^2 + Mx + N + \left(40a^2x^2 - 80ax + 40\right)K\sqrt{3}\arctan\left[\frac{1}{9}(14ax-15)\sqrt{3}\right]}{36a^2c^2(ax-1)^2}} \tag{38}$$

$$\frac{\Delta}{c} = 4x(1-ax)^3 \frac{\ddot{y}}{y} - 3a(1-ax)^2\left(1 + 2x\frac{\dot{y}}{y}\right) + 3a - 3a^2x + a^3x^2 \tag{39}$$

## 4. Physical Requirements for the New Model

For a model to be physically acceptable, the following conditions should be satisfied [4,49]:

(i) The metric potentials $e^{2\lambda}$ and $e^{2\nu}$ assume finite values throughout the stellar interior and are singularity-free at the center $r=0$.

(ii) The energy density $\rho$ should be positive and a decreasing function inside the star.

(iii) The radial pressure also should be positive and a decreasing function of radial parameter.

(iv) The radial pressure and density gradients $dp_r/dr \leq 0$ and $d\rho/dr \leq 0$ for $0 \leq r \leq R$.

(v) The anisotropy is zero at the center $r=0$, i.e. $\Delta(r=0) = 0$.

(vi) Any physically acceptable model must satisfy the causality condition, that is, for the radial sound speed $v_{sr}^2 = \frac{dp_r}{d\rho}$, we should have $0 \leq v_{sr}^2 \leq 1$.

(vii) The interior solution should match with the Schwarzschild exterior solution, for which the metric is given by

$$ds^2 = -\left(1 - \frac{2M}{r}\right)dt^2 + \left(1 - \frac{2M}{r}\right)^{-1} dr^2 + r^2(d\theta^2 + \sin^2\theta d\varphi^2) \tag{40}$$

through the boundary $r=R$ where $M$ is the total mass of the sphere

The conditions (ii) and (iv) imply that the energy density must reach a maximum at the centre and decreasing towards the surface of the sphere.

## 5. Physical Analysis

With $n=2$, the metric potentials $e^{2\lambda}$ and $e^{2\nu}$ have finite values and remain positive throughout the stellar interior. At the center $e^{2\lambda(0)}=1$ and $e^{2\nu(0)} = A^2 c_1^2 (-6)^{2A^*}(-1)^{2B} e^{-\frac{E}{2}}$. We show that in $r=0$ $\left(e^{2\lambda(r)}\right)'_{r=0} = \left(e^{2\nu(r)}\right)'_{r=0} = 0$ and this makes is possible to verify that the gravitational potentials are regular at the center.

The energy density and radial pressure are positive and well behaved between the center and the surface of the star. In the center $\rho(r=0)=6ac$ and $p_r(r=0)=36Ha^2c^2 - \dfrac{K}{6ac}$, therefore the energy density will be non-negative in $r=0$ and $p_r(r=0) > 0$. In the surface of the star $r=R$ and we have $p_r(r=R)=0$ and $R = \sqrt{\dfrac{1}{5ac}\left(6-\left(\dfrac{K}{Ha^3 c^3}\right)^{1/3}\right)}$. For the radial pressure of density gradients we obtain

$$\frac{d\rho}{dr} = -10 a^2 c^2 r \tag{41}$$

$$\frac{dp_r}{dr} = -20 Ha^3 c^3 (6-5acr^2) r - \frac{10Kr}{(6-5acr^2)^2} \tag{42}$$

The energy density and radial pressure decrease from the centre to the surface of the star. From eq.(14), the mass function can be written as

$$M(r) = \frac{1}{2} acr^3 (2 - acr^3) \tag{43}$$

and the total mass of the star is

$$M(r=R) = \frac{1}{2} ac \left[\frac{1}{5ac}\left(6-\left(\frac{K}{Ha^3 c^3}\right)^{1/3}\right)\right]^{3/2} \left\{2 - ac\left[\frac{1}{5ac}\left(6-\left(\frac{K}{Ha^3 c^3}\right)^{1/3}\right)\right]\right\} \tag{44}$$

To maintain of causality, the radial sound speed defined as $v_{sr}^2 = \frac{dp_r}{d\rho}$, should be within the limit $0 \leq v_{sr}^2 \leq 1$ in the interior of the star. For this model

$$v_{sr}^2 = \frac{dp_r}{d\rho} = \frac{2Ha^3c^3(6-5ax)^3 + K}{a^2c^2(6-5ax)^2} \qquad (45)$$

and for the equation (45), we can impose the condition

$$0 \leq \frac{2Ha^3c^3(6-5acr^2)^3 + K}{a^2c^2(6-5acr^2)^2} \leq 1 \qquad (46)$$

Matching conditions for $r=R$ can be written as

$$\left(1 - \frac{2M}{R}\right)^{-1} = \frac{1}{(1-acR^2)^2} \quad \text{and} \qquad (47)$$

$$\left(1 - \frac{2M}{R}\right) = A^2 y^2(cr^2) \qquad (48)$$

In the Table I shows the values of the parameters $a$, $\alpha$, $\beta$, $H$ and $K$ of the energy density and radial pressure for $n=2$.

**Table I.** Parameters $a$, $\alpha$, $\beta$, $H$ and $K$ for $n=2$.

| a | α | β | H | K(x10⁻⁶) |
|---|---|---|---|---|
| 0.10 | 2 | 1 | 0.5 | 5 |
| 0.11 | 2 | 1 | 0.5 | 5 |
| 0.12 | 2 | 1 | 0.5 | 5 |

The figures 1, 2, 3, 4, 5, 6 and 7 present the dependence of $\rho$, $\frac{d\rho}{dr}$, $p_r$, $\frac{dp_r}{dr}$, $M$, $\Delta$ and $v_{sr}^2$ with the radial coordinate for the parameters given in the Table I.

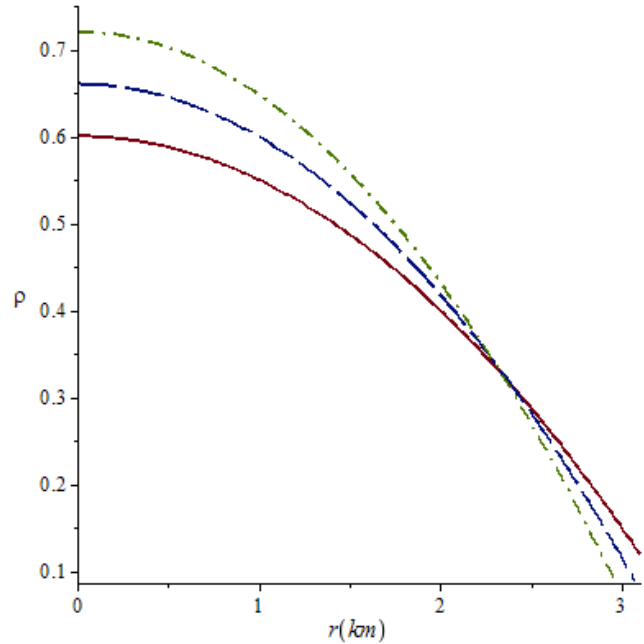

*Figure 1.* Energy density against radial coordinate for the parameters given in Table I. It has been considered that *a=0.1* (solid line) ; *a=0.11* (long-dash line); *a=0.12* (dashdot line).

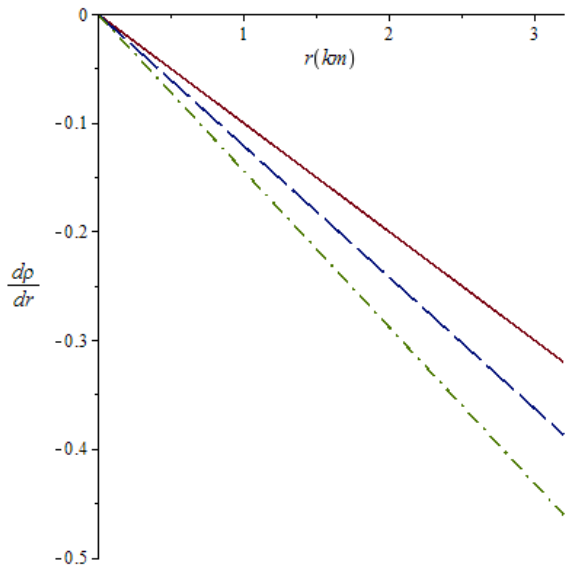

*Figure 2.* Density gradient against radial coordinate for the parameters given in Table I. It has been considered that *a=0.1* (solid line) ; *a=0.11* (long-dash line); *a=0.12* (dashdot line).

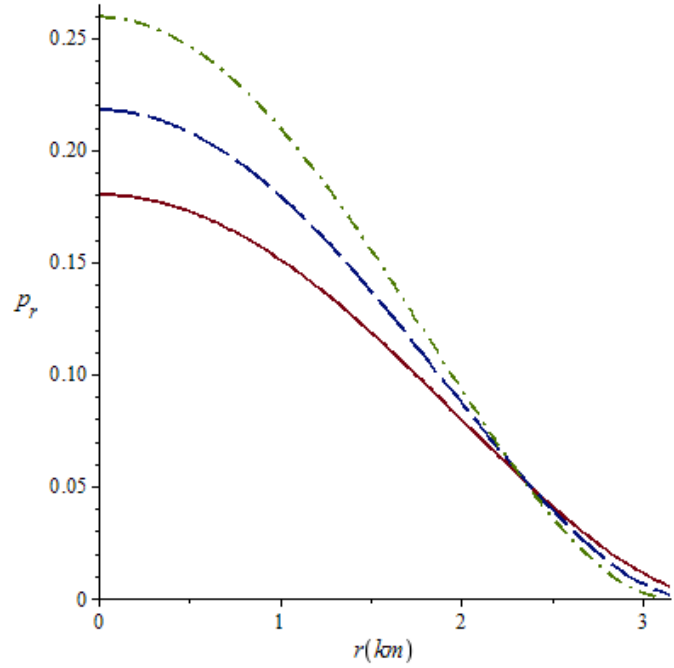

*Figure 3.* Radial pressure against radial coordinate for the parameters given in Table I. It has been considered that *a=0.1* (solid line) ; *a=0.11* (long-dash line); *a=0.12* (dashdot line).

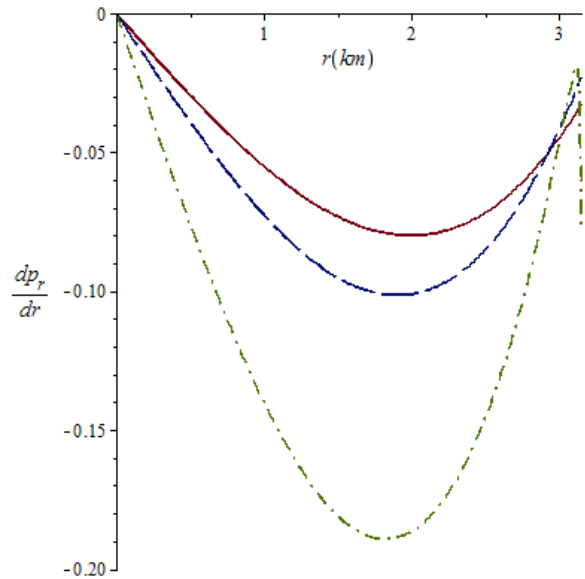

*Figure 4.* Radial pressure gradient against radial coordinate for the parameters given in Table I. It has been considered that *a=0.1* (solid line) ; *a=0.11* (long-dash line); *a=0.12* (dashdot line).

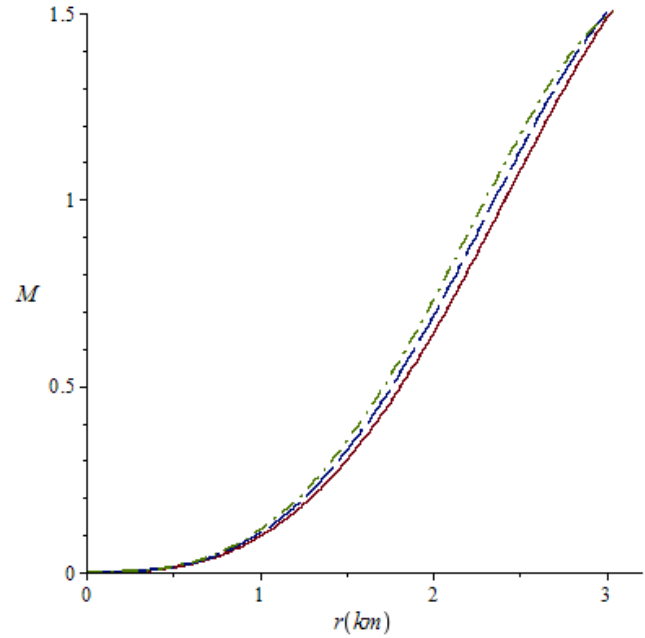

*Figure 5.* Mass function against radial parameter for the parameters given in Table I. It has been considered that *a=0.1* (solid line) ; *a=0.11* (long-dash line); *a=0.12* (dashdot line).

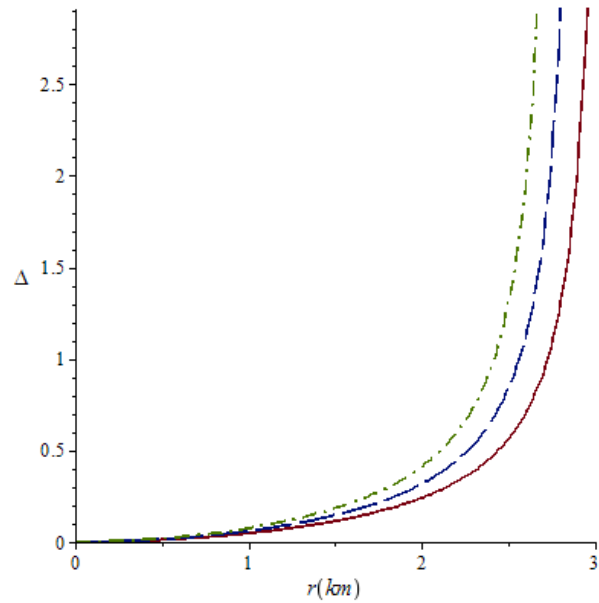

*Figure 6.* Anisotropy against radial parameter for the parameters given in Table I. It has been considered that *a=0.1* (solid line) ; *a=0.11* (long-dash line); *a=0.12* (dashdot line).

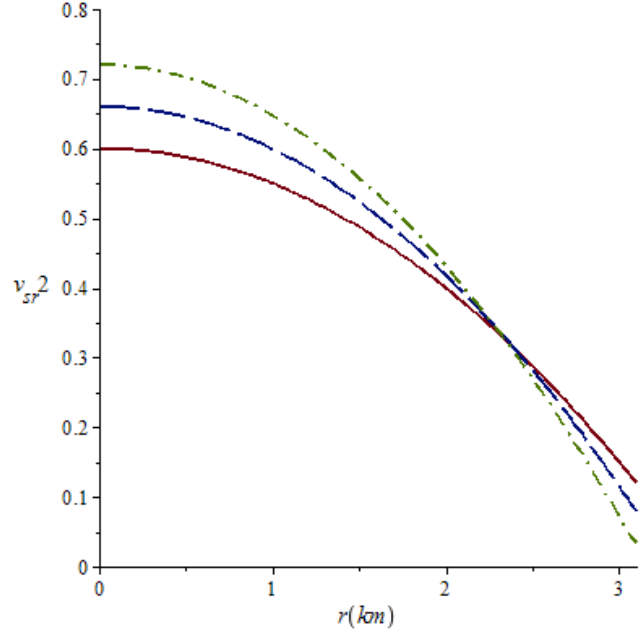

*Figure 7.* Radial speed sound against radial parameter for the parameters given in Table I. It has been considered that *a=0.1* (solid line) ; *a=0.11* (long-dash line); *a=0.12* (dashdot line).

In the Figure 1 is shown that the energy density remains positive, continuous and is monotonically decreasing function throughout the stellar interior for all values of a. In the Figure 2 it is noted that for the radial variation of energy density gradient $\frac{d\rho}{dr}<0$ in the three cases studied. The radial pressure showed the same behavior by the energy density, that is, it is growing within the star and vanishes at a greater radial distance, but takes the higher values when *a* is increased and its results are shown in Figure 3. Again, according to Figure 4, the profile of $\frac{dp_r}{dr}$ shows that radial pressure gradient is negative inside the star. In Figure 5, the mass function is continuous, strictly increasing and well behaved for all the cases. The anisotropic factor is plotted in Figure 6 and it shows that vanishes at the centre of the star, i.e. Δ(*r=0*) =0 [42,49]. We can also note that Δ admits higher values when *a* increases.

A physically acceptable model must satisfy the causality condition, i.e., the radial sound speed must be within the range $0 \leq v_{sr}^2 \leq 1$ [4,44]. The profile of radial speed sound is plotted in Figure 7 for different values of *a*. In all the cases $v_{sr}^2$ is in the expected range and is a monotonic decreasing function with the radial parameter.

For the case n=3, we have for the metric potentials $e^{2\lambda(0)}=1$,
$e^{2\nu(0)} = A^2 c_2^2 (9)^{2F} (-1)^{2G} e^{-\frac{N+40K\sqrt{3}\arctan\left[\frac{-15\sqrt{3}}{9}\right]}{36a^2c^2}}$ and $\left(e^{2\lambda(r)}\right)'_{r=0} = \left(e^{2\nu(r)}\right)'_{r=0} = 0$ at the centre r=0. Again the gravitational potentials are regular in the origin.

The energy density and radial pressure also are positive and well behaved in the stellar interior. In the center $\rho(r=0) = 9ac$ and $p_r(r=0) = 81Ha^2c^2 - \frac{K}{9ac}$, therefore the energy density will be non-negative in r=0 and $p_r(r=0) > 0$. In the surface of the star $r=R$ and we have $p_r(r=R)=0$ and $R = \frac{\sqrt{210ac - 14\sqrt{28ac\left(\frac{K}{H}\right)^{1/3} - 27a^2c^2}}}{14ac}$. For the radial pressure of density gradients we obtain

$$\frac{d\rho}{dr} = 2a^2c^2 r\left(-15r + 14acr^3\right) \tag{49}$$

$$\frac{dp_r}{dr} = 2Ha^2c^2\left(7a^2c^2r^4 - 15acr^2 + 9\right)\left(28a^2c^2r^3 - 30acr\right) + \frac{K\left(28a^2c^2r^3 - 30acr\right)}{ac\left(7a^2c^2r^4 - 15acr^2 + 9\right)^2} \tag{50}$$

Then according to the equations (49) and (50), the energy density and radial pressure decrease from the centre to the surface of the star. For the mass function we have from equation (27)

$$M(r) = \frac{1}{2}acr^3\left(3 - 3acr^2 + a^2c^2r^4\right) \tag{51}$$

and the total mass of the star can be written as

$$M(r=R) = \frac{\left(210ac - 14\sqrt{28ac\left(\frac{K}{H}\right)^{1/3} - 27a^2c^2}\right)^{3/2}}{5488a^2c^2} \left[ 3 - 3\left(\frac{210ac - 14\sqrt{28ac\left(\frac{H}{K}\right)^{1/3} - 27a^2c^2}}{196ac}\right) + \frac{\left(210ac - 14\sqrt{28ac\left(\frac{H}{K}\right)^{1/3} - 27a^2c^2}\right)^2}{38416a^2c^2} \right]$$

(52)

Again to maintain causality the condition $0 \leq v_{sr}^2 \leq 1$ implies that

$$0 \leq \frac{2Ha^3c^3\left(9 - 15acr^2 + 7a^2c^2r^4\right)^3 + K}{a^2c^2\left(9 - 15acr^2 + 7a^2c^2r^4\right)^2} \leq 1$$

(53)

And for the Matching conditions across the boundary $r=R$ we have

$$\left(1 - \frac{2M}{R}\right)^{-1} = \frac{1}{\left(1 - acR^2\right)^3} \quad \text{and} \quad \left(1 - \frac{2M}{R}\right) = A^2 y^2\left(cr^2\right)$$

In the Table II shows the values of the parameters $a$, $\alpha$, $\beta$, $H$ and $K$ of the energy density and radial pressure for $n=3$.

**Table II**. Parameters $a$, $\alpha$, $\beta$, $H$ and $K$ for $n=3$.

| $a$ | $\alpha$ | $\beta$ | $H$ | $K(x10^{-3})$ |
|---|---|---|---|---|
| 0.10 | 2 | 1 | 0.2 | 5 |
| 0.11 | 2 | 1 | 0.2 | 5 |
| 0.12 | 2 | 1 | 0.2 | 5 |

The figures 8, 9, 10, 11, 12, 13 and 14 present the dependence of $\rho$, $\frac{d\rho}{dr}$, $p_r$, $\frac{dp_r}{dr}$, $M$, $\Delta$ and $v_{sr}^2$ with the radial coordinate for the parameters given in the Table II.

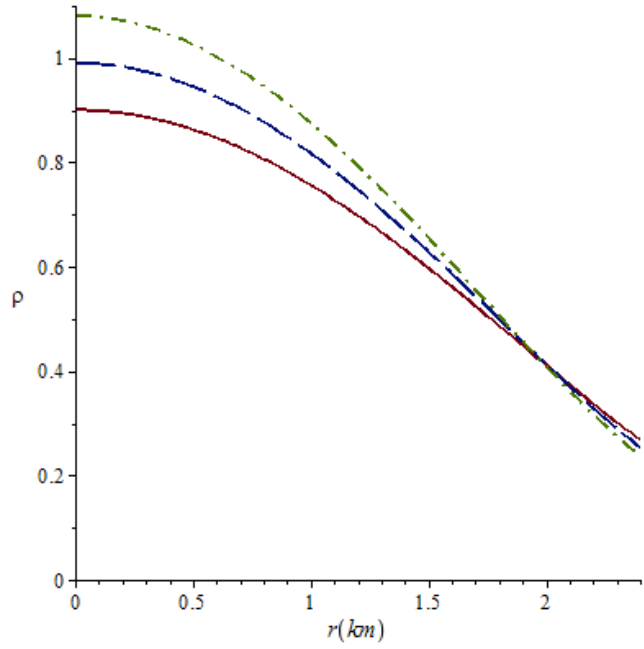

*Figure 8.* Energy density against radial coordinate for the parameters given in Table II. It has been considered that *a=0.1* (solid line); *a=0.11* (long-dash line); *a=0.12* (dashdot line).

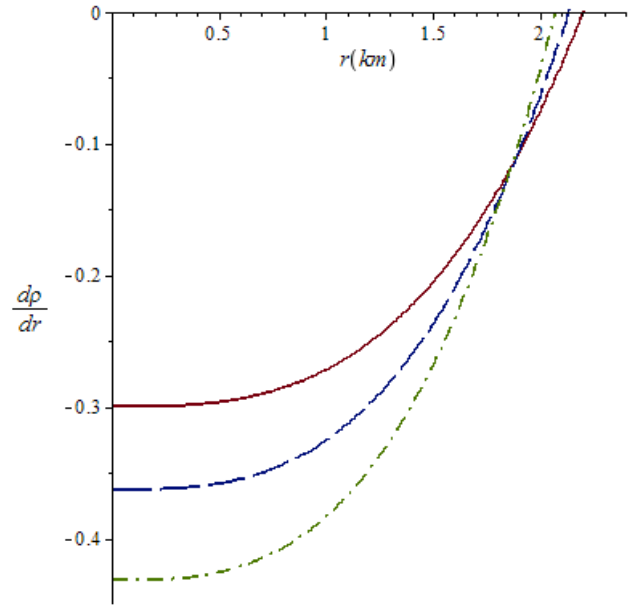

*Figure 9.* Density gradient against radial coordinate for the parameters given in Table II. It has been considered that *a=0.1* (solid line) ; *a=0.11* (long-dash line); *a=0.12* (dashdot line).

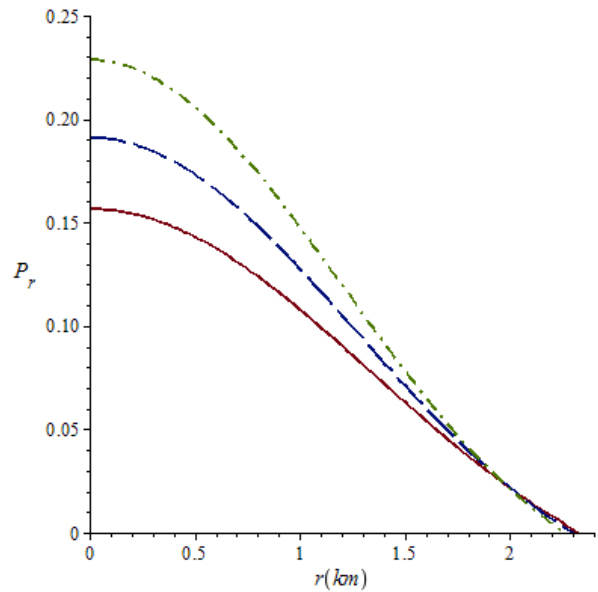

*Figure 10.* Radial pressure against radial coordinate for the parameters given in Table II. It has been considered that *a=0.1* (solid line) ; *a=0.11* (long-dash line); *a=0.12* (dashdot line).

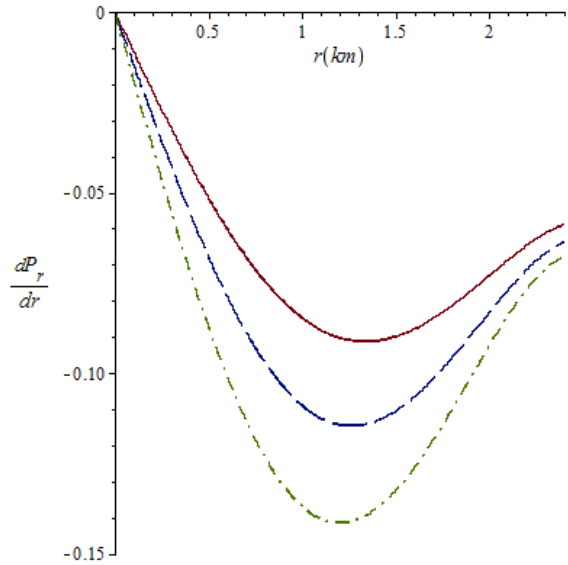

*Figure 11.* Radial pressure gradient against radial coordinate for the parameters given in Table II. It has been considered that $a=0.1$ (solid line); $a=0.11$ (long-dash line); $a=0.12$ (dashdot line).

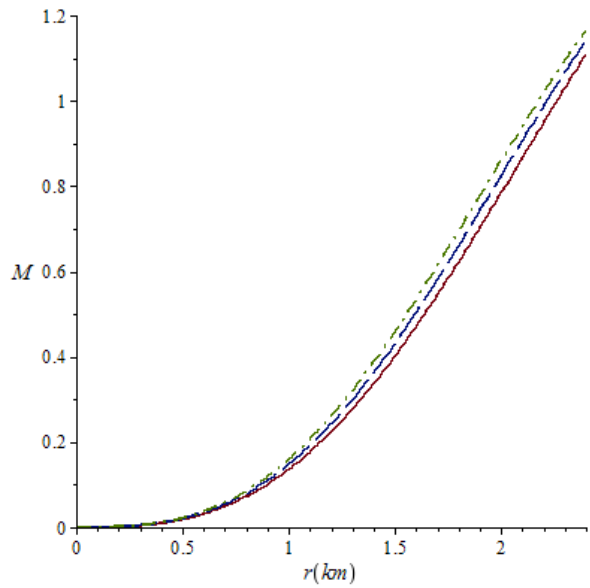

*Figure 12.* Mass function against radial parameter for the parameters given in Table II. It has been considered that $a=0.1$ (solid line); $a=0.11$ (long-dash line); $a=0.12$ (dashdot line).

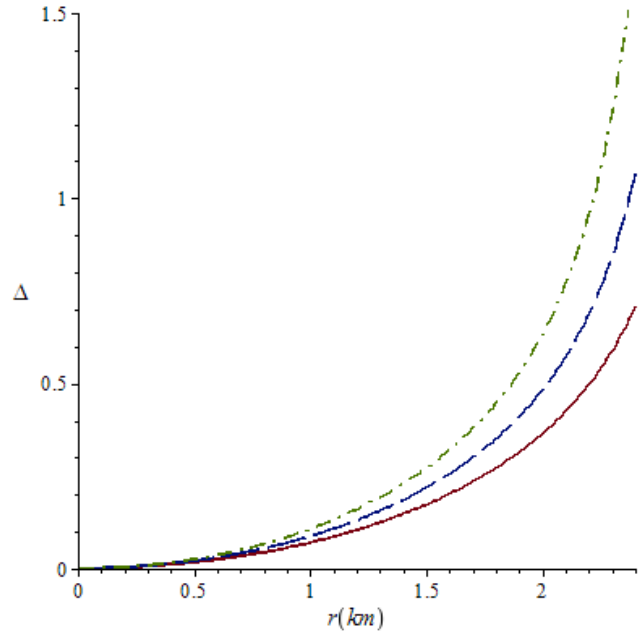

*Figure 13.* Anisotropy against radial parameter for the parameters given in Table II. It has been considered that *a=0.1* (solid line); *a=0.11* (long-dash line); *a=0.12* (dashdot line).

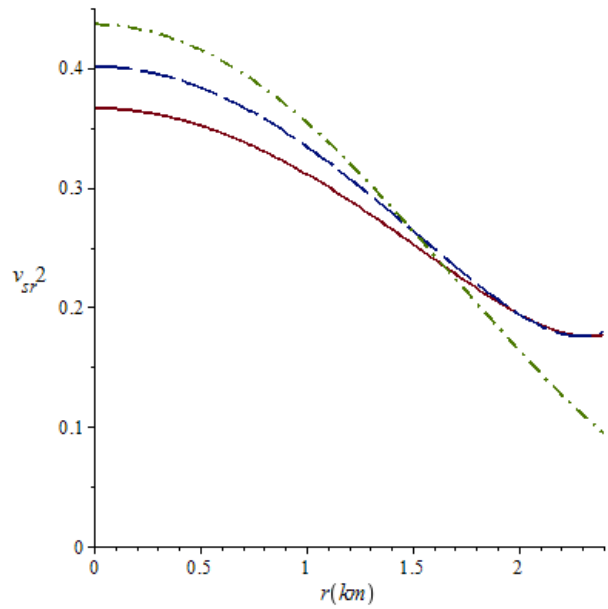

*Figure 14.* Radial speed sound against radial parameter for the parameters given in Table II. It has been considered that *a=0.1* (solid line); *a=0.11* (long-dash line); *a=0.12* (dashdot line).

As in the case *n=2*, the energy density is continuous and is monotonically decreasing inside the star and increases with values of a (Figure 8) and the radial variation of energy density gradient has been shown in Figure 9, in which it is also observed that $\frac{d\rho}{dr}<0$. With *n=3*, the radial pressure always is positive throughout the stellar interior and vanishes at a finite radial distance and its results are shown in Figure 10. Again, the radial pressure increases when *a* takes higher values. In the Figure 11, it is also verified that the gradient $\frac{dp_r}{dr}$ is negative inside the star. The Figure 12 shows that for *n=3* the mass function is regular, well behaved and strictly increasing from the centre to the surface of the star. In the Figure 13, the anisotropy ∆ also is zero at the center *r=0* and its value increase when a takes higher values. Again, as shown in Figure 14, the radial speed of sound is always less than the unity and the causality condition is maintained in the stellar interior and it is important physical requirement as indicated Delgaty and Lake [4].

We can compare the values calculated for the mass function with some experimental results. For *n=2* the values of a allow to obtain the mass $1.69 M_\odot$ which can correspond to the astronomic object X-ray pulsar Her X-1. With *n=3*, the obtained results with the values of a selected produce the mass $1.2 M_\odot$ which can be associated with the compact star RX J1856-37.

## 6. Conclusions

In this paper we have generated new models of anisotropic stars considering the Thirukkanesh-Ragel-Malaver ansatz for the gravitational potential and modified version of Chaplygin equation of state proposed for Errehymy and Daoud of the form $p_r = H\rho^\alpha - K\rho^{-\beta}$ [30]. These models may be used in the description of compact objects in absence of charge and in the study of internal structure of strange quark stars. We show that the developed configuration obeys the physical conditions required for the physical viability of the stellar model. A graphical analysis shows that the radial pressure, energy density, mass function and anisotropy are regular at the origin and well behaved in the interior. The new solutions match smoothly with the Schwarzschild exterior metric at the boundary *r=R because* matter variables and the gravitational potentials of this work are consistent with the physical analysis of these stars. It is expected that the results of this research can contribute to modeling of relativistic compact objects and configurations with anisotropic matter distribution.

The obtained models can be related to astronomical objects RX J1856-37, X-ray pulsar Her X-1, 4U 1728-34 and SAX J1808.4-3658. physical features associated with the matter, radial pressure, density, anisotropy, gravitational potential, and the graphical approach suggests that the model for the star RX J1856 - 37 is well behaved. It is important to mention that a strange star model may be compatible with the compact star SAX J1808.4-3658.

Compact stars typically may undergo vortex action fields to churn energy and matter. Breakthrough formalism examining quantum fields point model algorithmically gaged to electromagnetic fields provided string metrics that are associated quantum to mesoscopic to astrophysics; hold Plenum PDP circuit assemblages helped to identify mechanism generator of sustainable energy at quantum to astrophysical levels, having vacuum friction causing the inertial matter linking to non-inertial vacuum.

Quantum astrophysical signal/noise density matrix with prime factorized magic square symmetry mechanistic processes have powerful methodology to characterize compact, anisotropic, strange Star systems that essentially balance gas clouds such as Chaplygin equation of state to hydrogenous energetics within vacuum multiverse. We believe that critical quantum signal/noise density matrix values may input to astrophysics physical features associating the matter, radial pressure, density, anisotropy, gravitational potential, and energy density Schwarzschild-Einstein-Maxwell metrics.

## Conflicts of Interest
The authors declare that there is no conflict of interest regarding the publication of this article.

## Funding
This research received no specific grant from any funding agency in the public, commercial or not-for-profit sectors.